\documentclass[aps,prl,preprint,showpacs]{revtex4}
\usepackage{dcolumn} 
\usepackage{bm}% bold math
\usepackage{latexsym}
\usepackage{color,graphicx}
\begin{document}
% \draft

\thispagestyle{empty}

{\baselineskip0pt
\leftline{\large\baselineskip16pt\sl\vbox to0pt{\hbox{\it Department of
Mathematics and Physics}
               \hbox{\it Osaka City  University}\vss}}
\rightline{\large\baselineskip16pt\rm\vbox to20pt{\hbox{OCU-PHYS-334}
            \hbox{AP-GR-79}
\vss}}%
}
\vskip3cm

\title{Visible borders of spacetime generated by high-energy collisions}
\author{
Ken-ichi Nakao$^{1}$\footnote{Electronic
address: knakao@sci.osaka-cu.ac.jp}, 
Tomohiro Harada$^{2}$\footnote{Electronic
address: harada@rikkyo.ac.jp}  
and 
Umpei Miyamoto$^{2}$\footnote{Electronic
address: umpei@rikkyo.ac.jp} 
}
\affiliation{
$^{1}$Department of Physics, Graduate School of Science, Osaka City University, 
Osaka 558-8585, Japan \\
$^{2}$Department of Physics, Rikkyo University, Toshima, 
Tokyo 171-8501, Japan.
}
\date{\today}

\begin{abstract}                % DON'T CHANGE THIS LINE
Several years ago, two of the present 
authors proposed 
the concept of the border of spacetime as a 
generalization of spacetime singularities. Visible borders of spacetime, 
which replace naked singularities of classical theory, are not only necessary for the 
mathematical completeness of general relativity
but they also provide a window into new physics of strongly curved
spacetime, which is observable in principle.
By employing simple geometrical and dimensional arguments, we show that 
not only black holes but also visible borders of spacetime 
will be generated at, for example, the CERN Large Hadron Collider, 
provided that the energy scale of quantum gravity is near 1 TeV 
in the framework of the large extra-dimension scenario. 
\end{abstract}

\pacs{04.20.Dw, 04.20.Cv, 11.10.Kk, 04.62.+v}

\maketitle
%%---------------------------%

In the framework of the large extra-dimension scenario, 
the fundamental Planck scale $E_P$ can be lowered to a terra-electron-volt (TeV) scale 
to resolve the hierarchy problem between the four-dimensional 
Planck scale ($10^{19}$ GeV) and the electroweak scale
($100$ GeV)~\cite{Arkani-hamed,RS}.
One of the exciting predictions of this scenario is that 
substantial black holes may be generated at the CERN Large Hadron Collider (LHC) 
and identified through Hawking radiation~\cite{GT01,DL01,Ida_etal}. 
Black holes are one key to solving problems related to the quantum 
nature of gravity.
Hence, the observation of black holes 
at LHC would represent the first experimental study of quantum gravity. 

However, here we consider whether quantum effects 
of gravity can be observed only through black holes. Within the 
framework of classical general relativity, singularity theorems 
predict that development from generic 
initial data results in singular spacetime~\cite{Wald}. 
Although these theorems do not specify the nature of spacetime singularities 
(beside their geodesic incompleteness),
these singularities will be accompanied by 
divergences in the energy density, 
stress and spacetime curvature (i.e., tidal forces), etc. 
The big bang is the most well-known example of spacetime singularities. 
Since all the known laws of physics (including general relativity itself) 
will lose their validity at spacetime singularities, 
it is not possible to predict what happens there without knowledge of new physics such as 
a quantum theory of gravity. 
A spacetime singularity that can causally affect regular regions is termed 
a naked singularity. If naked singularities
exist, it should, in principle, be possible to observe something from 
these singularities. The big bang can be regarded as a naked 
singularity and hence if we obtain any direct signals from it, 
it will be possible to learn something about the quantum theory of gravity. 
If naked singularities besides the big bang exist in our universe,
they might also be arenas of quantum gravity.

Penrose proposed a conjecture, known as the cosmic censorship hypothesis, 
regarding the visibility of spacetime singularities~\cite{penrose1969}. 
There are two versions of this hypothesis. 
For spacetimes that contain physically reasonable matter fields 
and that are developed from generic nonsingular initial data, the weak version of 
the hypothesis claims that 
no spacetime singularity is visible from infinity, whereas the strong version 
claims that the spacetime is globally hyperbolic, or roughly speaking, 
that there is no unpredictable region in the future of the initial data. 
A spacetime singularity that conflicts with the weak version is called 
a globally naked singularity, whereas a naked singularity that does not conflict 
with the weak version is called a locally naked singularity. 
This hypothesis plays a crucial role in the proofs of 
theorems for black holes and of singularity theorems. 
If it is true, it will be impossible to observe 
spacetime singularities besides the big bang. 
However, several candidates for counterexamples to 
the cosmic censorship hypothesis has been theoretically found~\cite{Joshi07}. 
Charged black holes and rotating black holes are 
regarded as serious counterexamples to the strong version of the hypothesis since 
there are unpredictable regions within charged or rotating 
black holes, although the existence of such unpredictable regions 
is not necessarily related to naked singularities~\cite{Brady,Burko,Dafermos,Ori}.

It seems likely that spacetime singularities are accompanied 
by very strong gravity and thus are regarded as fatal disasters. 
However, this is not always true. 
Spacetime singularities with strong gravity 
are surrounded by a domain from which even light rays cannot escape. 
Such singularities are enclosed by horizons and thus cannot be 
counterexamples to the weak version of the cosmic censorship hypothesis. 
In the case of collapse of matter fields due to 
their self-gravity, which is so strong that it overcomes their own stress, 
the resultant singularities are not expected to be globally naked. 
In other words, a naked singularity can be globally naked due to the 
weakness of its gravity, which is too weak to trap light rays within 
the vicinity of the singularity. If this is true, the formation of globally naked singularities 
requires matter that has stress that is too weak to interrupt the collapse 
caused by the relatively weak self-gravity. Thus, in 
the cosmic censorship formulated by Wald, {\it physically natural matter 
fields} should not form 
singularities if there is no gravity~\cite{Wald,Wald97}. 
In fact, all known candidates for counterexamples to the weak cosmic censorship 
hypothesis involve matter fields that do not satisfy Wald's condition or 
have not been shown to develop from generic nonsingular initial data. 

The cosmic censorship hypothesis is formulated 
within the framework of general relativity. However, since general relativity 
is a classical theory, it becomes invalid at energy scales higher than 
the fundamental Planck scale $E_P$. 
Thus, it is more practical to regard a region 
in which general relativity loses its validity as an effective spacetime singularity. 
Based on this point of view, Harada and Nakao proposed the concept of the border 
of spacetime~\cite{HN2004}. 
The domain ${\cal A}$ of spacetime is called a border if 
the following inequality is satisfied:
\begin{equation}
\inf_{\cal A} F \geq \alpha E_P^2,  \label{criterion}
\end{equation}
where $\alpha$ is a positive constant of order unity and $F$ is given, for instance, by 
using the scalar polynomials of the Riemann tensor, 
or components of the Riemann tensor with respect to a parallelly transported 
orthonormal basis $e_\mu^a$ as 
\begin{equation}
	F
	:=
	{\rm max}\left(|R^a_a|,|R^{ab}R_{ab}|^{1/2},|R^{abcd}R_{abcd}|^{1/2},
	|R_{\mu\nu\rho\sigma}|\right).
\label{eq:strength}
\end{equation}
(Hereafter we adopt the natural units $c=\hbar=1$ and the 
abstract index notation: Latin indices denote the 
tensor type and Greek indices denote the component 
with respect to some basis vectors~\cite{Wald}.)
We denote the union of all borders in spacetime 
${\cal M}$ by ${\cal U}_{\rm B}$. 
We call a border ${\cal A}$ a visible border if 
$J^{+}({\cal A},{\cal M})\cup ({\cal M}-{\cal U}_{\rm B})$ is not empty, 
where $J^{+}({\cal A},{\cal M})$ is the causal future of ${\cal A}$ 
in ${\cal M}$. 
We can also naturally introduce a globally visible border as an extension 
of a globally naked singularity~\cite{HN2004}.

It is instructive to consider 
cylindrically symmetric gravitational collapse in four-dimensional spacetime 
that is asymptotically flat in the spacelike direction orthogonal 
to the symmetry axis. No horizon forms in this system provided the stress-energy 
tensor $T_{ab}$ satisfies 
$T_{ab}k^ak^b\geq0$ for any null vector $k^a$ 
~\cite{Hayward}. Thus, if a spacetime 
singularity forms, it is necessarily naked. 
Several studies of this cylindrical 
system suggest that it is probable that matter fields with positive tangential 
stresses that satisfy physically reasonable equations 
of state do not form a spacetime singularity 
in gravitational collapse since the tangential stress will finally 
overcome the cylindrical 
gravity and, as a result, a bounce will occur~\cite{Piran78,AT92,NM05,NIK08}. 
On the other hand, if the initial imploding velocity is sufficiently large, 
the energy density and stress 
in the vicinity of the symmetry axis can be momentarily very large. 
In addition, $F$ defined by Eq.~(\ref{eq:strength}) 
can also become very large through the Einstein equations. 
Thus, in such a situation, 
the neighborhood of the symmetry axis can be a globally visible 
border of spacetime, even though the spacetime remains regular. 

Giddings and Thomas~\cite{GT01} and 
Dimopoulos and Landsberg~\cite{DL01} independently 
pointed out that if the TeV scale gravity scenario is true, 
since gravitational attraction in a short range 
is much stronger than that in a long range, 
collisions of protons will generate so many black holes 
at the LHC that the signals of their production should be 
observable (an excellent review of this is given in Ref.~\cite{Kanti}). 
At the high-energy collider, particles are accelerated and forced 
to collide with each other. In other words, a very high energy density 
region will appear due to high initial velocities 
and, in this sense, 
the situation is similar to the formation of 
a cylindrical border mentioned above. Thus, not only black holes 
but also globally visible borders might form at the LHC. 
We are investigating
whether a region with a very high energy density appears without 
black hole formation.

In this paper, we consider a collision of two elementary particles. 
For simplicity, we assume that both particles are identical and structureless. 
The extra dimensions are assumed to be compactified 
into a length scale considerably larger than the fundamental 
Planck length $E_P^{-1}$, whereas the energy of the particles is assumed 
to be confined within the length scale $E_P^{-1}$ 
in every direction of the extra dimensions; i.e., within a 
volume $V_{D-4} E_P^{-(D-4)}$ in the extra dimensions, 
where $V_n$ is the volume of a unit $n$-ball
\begin{equation}
V_n=\pi^{\frac{n}{2}}/\Gamma\left(\frac{n}{2}+1\right),
\end{equation}
where $\Gamma(x)$ is the gamma function. 
Then, we consider the distribution of two particles in the remaining 
three-dimensional space.

We denote the center-of-mass (CM) energy of the two particles by $E$, which is much larger 
than their rest masses, and 
assume that the three momenta of these particles are parallel to each other. 
The minimum uncertainty in the position of each particle measured in the longitudinal 
direction of their three momenta is approximately $E^{-1}$ due to 
the Lorentz contraction.

In Refs.~\cite{GT01} and \cite{DL01}, it is assumed that 
these two particles will form a black hole if they pass through a transverse 
disk of radius $R_{\rm g}$ at almost 
the same time,
where $R_{\rm g}$ is the horizon radius of the Schwarzschild--Tangherlini 
black hole with gravitational mass $E$, 
\begin{equation}
R_{\rm g}:=E_P^{-1}\left(\frac{E}{E_P}\right)^{\frac{1}{D-3}}, 
\end{equation}
and where we have defined the fundamental Planck scale 
$E_P$ so that the reduced Compton wavelength agrees with the
gravitational radius if the particle mass is equal to $E_P$. 
In this normalization, $D$-dimensional 
Einstein's gravitational constant $\kappa$ ($G_{ab}=\kappa T_{ab}$) 
is related to $E_P$ in the form 
\begin{equation}
\kappa = \frac{1}{2}(D-2)S_{D-2}E_P^{2-D},
\end{equation}
where $S_n=(n+1)V_{n+1}$ is the area of a unit $n$-sphere. 
Based on this picture, the total cross section of
black hole formation is given by
\begin{equation} 
\sigma_{\rm tot}=\pi R_{\rm g}^2. \label{c-section}
\end{equation}

We consider the case in which these two particles 
pass through a transverse disk with a radius of 
$\nu R_{\rm g}$, where $\nu >1$. In this case, a black hole does not form. 
When these two particles pass through a disk of radius $\nu R_{\rm g}$ at 
almost the same time, they can be momentarily confined within a 
$(D-1)$-volume $V\simeq\pi \nu^2R_{\rm g}^2\times E^{-1}\times V_{D-4} E_P^{-(D-4)}$, 
where the second factor $E^{-1}$ is the uncertainty in the longitudinal position 
of each particle and $ V_{D-4} E_P^{-(D-4)} $ is the volume in the extra dimensions. 
At this moment, the average energy density $\rho$ in this volume is estimated to be
\begin{equation}
	\rho\simeq \frac{E}{V}
	=
	\frac{E_P^D}{\pi \nu^2 V_{D-4} }
	\left(
		\frac{E}{E_P}
	\right)^{ 2\left(  \frac{D-4}{D-3}  \right) }.
\label{rho}
\end{equation}
If $\kappa\rho$ is larger than or equal to $\alpha E_P^2$, then, 
through the Einstein equation, the average Ricci tensor satisfies 
the criterion (\ref{criterion}) for a border. 
The average density $\rho$ 
is a monotonically increasing function of the CM energy $E$. 
Thus, a sufficiently large CM energy can produce a visible border, since 
no black hole forms in this situation. 
The condition for realizing a border, $\kappa\rho \geq \alpha E_P^2$, 
and Eq.~(\ref{rho}) lead to 
\begin{equation}
	\nu^2
	<
	\nu_{\rm max}^2
	:=
        \frac{(D-2)S_{D-2}}{2\pi\alpha V_{D-4}}
%	\frac{1}{2\pi\alpha} 
%        \frac{\pi^{\frac{1}{2}}(D-1)\Gamma(D/2)}{\alpha\Gamma(D/2+1/2)}
	\left(
		\frac{E}{E_P}
	\right)^{	2\left( \frac{D-4}{D-3} \right)	}.
\label{numax}
\end{equation}
If the upper bound $\nu_{\rm max}^2$ is less than or equal to unity, the border will never
be visible. Since $\nu_{\rm max}^2$ is 
an increasing function of the CM energy, 
if $E> [ 2\pi\alpha (D-2)^{-1} S_{D-2}^{-1}
 V_{D-4}]^{\frac{1}{2}\left(\frac{D-3}{D-4}\right)}E_P$, then 
$\nu_{\rm max}^2$ is larger than unity, and the border will be visible.

%%%%%%%%%%%%%%%%%%%%%%%%%%%<<start table>>%%%%%%%%%%%%%%%%%%%%%%%%%%
%\begin{table}[htbp]
\begin{table}[h!]
%\footnotesize{
\caption{Values of $\alpha\nu_{\rm max}^2$ when $E=14$ TeV.}
\label{tab:nu2}
\setlength{\tabcolsep}{10pt}
%\begin{tabular}{|c|c|c|c|c|c|c|c|}
\begin{tabular}{cccccccc}
\hline\hline
$E_p$ & $D=5$ & $D=6$ & $D=7$ & $D=8$ & $D=9$ & $D=10$ & $D=11$
\\ \hline
%1 TeV & 66.0 & 180 & 309 & 437  & 1118  & 3372 & 783 
1 TeV & 66.0 & 180 & 309 & 437  & 559  & 674 & 783 
\\ \hline
%14 TeV & 4.71 & 5.33 & 5.90 & 6.40 & 13.7 & 36.6 & 7.73 
14 TeV & 4.71 & 5.33 & 5.89 & 6.40 & 6.87 & 7.31 & 7.73  
\\ \hline\hline
\end{tabular}
%}
\end{table}

The value of $(\nu_{\rm max}^2-1)$ expresses the ratio of the production rate of
visible borders to that of black holes.
The production rate of visible borders increases with increasing CM energy.
In the case of $E=14$ TeV which is equal to the CM energy achieved by the LHC, 
if the fundamental Planck scale $E_P$ is equal to 1 TeV, 
the production rate of visible borders will be about $70\alpha^{-1}$ times larger than 
the production rate of black holes for $D=5$, 
whereas it will about $800\alpha^{-1}$ times larger than that of black holes 
for $D=11$ (see Table \ref{tab:nu2}).
Although there is an ambiguous factor $\alpha$ 
in the criterion (\ref{criterion}) for the border, 
the present result, which is based on dimensional analysis, implies that 
if black holes form at the LHC, more visible borders will be produced than 
black holes.

$E$ will be larger than $E_P$ for the LHC if $E_P$ is about 1 TeV.
This implies that the minimum uncertainty $E^{-1}$ 
in the longitudinal position will be smaller than
the fundamental Planck length. If any structures smaller 
than the fundamental Planck length 
$E_P^{-1}$ are smeared out due to quantum gravitational 
effects, the uncertainty in the longitudinal 
direction should be 
estimated as the fundamental Planck length $E_P^{-1}$. 
If we adopt this assumption, the average energy density will be given by
$\rho \simeq E_P^D (E/E_P)^{\frac{D-5}{D-3}}/\pi\nu^2 V_{D-4}$. 
The corresponding $\nu_{\rm max}^2$ in this estimate is corrected only by
the factor $E_P/E$ compared with the estimate~(\ref{numax}).
Hence, for this case too, we may expect that visible borders will appear at the LHC
for $E \gtrsim E_P$. 
If it is necessary to account for quantum gravitational effects like these, 
we should regard such situations as situations in which the appearance of borders of 
spacetime is possible, even when the criterion (\ref{criterion}) is not satisfied. 

It is important to consider whether the visible border is gravitationally strong 
in the classical sense. In the framework of classical gravity (i.e., general relativity 
or Newtonian gravity), the codimension of singularity is closely
related 
to the strength of its gravity. 
In four-dimensional spacetime, the codimension of a point singularity is 
three, that of a line singularity is two, and that of a plane singularity is one. 
As is well known, a singularity of codimension three is fatal 
as an object of finite size will be crushed by its tidal 
force~\cite{tipler77,krolak87} if it has a 
non-zero gravitational mass. In the framework of general relativity, 
such singularities are enclosed by horizons. 
The tidal force of a codimension-two singularity is also 
strong enough to crush an object of finite size (except for conical 
singularities). However, since the gravity of a codimension-two singularity 
is generally weaker than that of a codimension-three singularity, 
it can be naked if it is sufficiently long and almost 
straight \cite{thorne1972,NST88}. 
A codimension-one singularity is gravitationally weak and thus is not fatal. 
Since visible borders produced by high-energy collisions are 
compressed in two or more directions 
(the longitudinal and bulk directions), they will be gravitationally strong 
in the classical sense. Thus, nonlinear quantum gravitational effects 
are expected to occur at such visible borders. 

For nongravitational interactions, the length scale of the physics that 
particle collisions reveal decreases with increasing CM energy.
However, this relation may not hold when gravity plays an important role.
If collisions on a super-Planckian energy scale necessarily 
produce only black holes (or in other words, super-Planckian phenomena 
necessarily appear through only black holes), higher CM energies will 
lead to long-range interactions rather than short-range interactions 
in the super-Planckian regime, because short-range interactions 
are hidden behind black holes and smeared out~\cite{GT01}. 
Probably, many researchers accept this picture. 
However, as we have seen in this paper, 
the larger CM energy of the particles may cause the formation of 
a visible border (i.e., a region with a super-Planckian 
average energy density that is
not hidden behind the event horizon). 
A larger average density corresponds to 
a shorter curvature radius through the Einstein equations. 
Thus, super-Planckian physics may lead to an extremely
short-range interaction through visible borders. 
At present, there is no precise prediction for what happens at the visible 
border since the theory of quantum gravity is not well developed.
However, if the gravity in the vicinity of the visible border is sufficiently 
strong and is well described by the classical theory of gravity, 
semiclassical particles will be created in curved spacetime. 
If the democratic nature of gravity 
still prevails there, all kinds of particles will be emitted from them 
through a similar mechanism 
as the emission of Hawking radiation from a black hole.
As mentioned above, if the visible border does not 
have any cutoff scales, the spectrum of emitted 
particles may be scale free, in contrast to the 
nearly black body radiation generated by Hawking evaporation of a 
black hole~\cite{Barve:1998ad,HIN,HINSTV01}.

\vskip0.5cm
\noindent
{\bf Acknowledgements}

KN is grateful to H.~Ishihara and his colleagues in the astrophysics and 
gravity group at Osaka City University for helpful 
discussions and criticism. KN and TH thank T.~Tanaka for very useful comments. 
TH and UM thank H. Tanaka for helpful comments. 
This work was partially supported by a Grant-in-Aid for Scientific Research
from the Ministry of Education, Culture, Sports, 
Science and Technology, Japan [Young Scientists (B) 21740190 (TH) and 22740176 (UM)].

\end{document}